\documentclass[showpacs, pra,twocolumn,preprintnumbers ,amsmath, amssymb, superscriptaddress, aps]{revtex4-2}
\usepackage{color}
\usepackage{amsmath,amssymb}
\usepackage{pifont}
\usepackage{amssymb}  
\usepackage{bbold}
\usepackage{float}
\usepackage{subfloat}
\usepackage{adjustbox}

\usepackage[caption=false]{subfig}
\usepackage{tikz}
\usepackage{makecell}
\usepackage{subfig}
\usepackage{pifont}   
\usepackage{graphicx} 
\graphicspath{{Figures/}}
\usepackage{dcolumn}  
\usepackage{bm}       
\usepackage{multirow} 
\usepackage{placeins}
\usepackage[colorlinks]{hyperref}
\usepackage{mathtools}
\usepackage{appendix}

\def \be{\begin{align}}
	\def \ee{\end{align}}
\def \bea{\begin{eqnarray}}
	\def \eea{\end{eqnarray}}

\begin{document}

        \title{Tunneling in ABC trilayer graphene superlattice}
        \date{\today}
        \author{Mouhamadou Hassane Saley }
        \affiliation{Laboratory of Theoretical Physics, Faculty of Sciences, Choua\"ib Doukkali University, PO Box 20, 24000 El Jadida, Morocco}
        \author{Jaouad El-hassouny }
        \affiliation{Laboratory of Theoretical Physics, Faculty of Sciences, Choua\"ib Doukkali University, PO Box 20, 24000 El Jadida, Morocco}
        \author{Abderrahim El Mouhafid}
        \affiliation{Laboratory of Theoretical Physics, Faculty of Sciences, Choua\"ib Doukkali University, PO Box 20, 24000 El Jadida, Morocco}
        \author{Ahmed Jellal}
        \email{a.jellal@ucd.ac.ma}
        \affiliation{Laboratory of Theoretical Physics, Faculty of Sciences, Choua\"ib Doukkali University, PO Box 20, 24000 El Jadida, Morocco}
        \affiliation{Canadian Quantum Research Center, 204-3002 32 Ave Vernon,  BC V1T 2L7, Canada}

       
        \begin{abstract}
        	
We study the transport properties of Dirac fermions in ABC trilayer graphene (ABC-TLG) superlattices. More specifically, we analyze the impact of varying the physical parameters—the number of cells, barrier/well width, and barrier heights—on electron tunneling in the ABC-TLG. In the initial stage, we solved the eigenvalue equation to determine the energy spectrum solutions for the ABC-TLG superlattices. Subsequently, we applied boundary conditions to the eigenspinors and employed the transfer matrix method to calculate transmission probabilities and conductance. For the two-band model, we identified the presence of Klein tunneling, with a notable decrease as the number of cells increased. The introduction of interlayer bias opened a gap as the number of cells increased, accompanied by an asymmetry in scattered transmission. Increasing the barrier/well width and the number of cells resulted in an amplified number of gaps and oscillations in both two-band and six-band cases. We observed a corresponding decrease in conductance as the number of cells increased, coinciding with the occurrence of a gap region. Our study demonstrates that manipulating parameters such as the number of cells, the width of the barrier/well, and the barrier heights provides a means of controlling electron tunneling and the occurrence of gaps in ABC-TLG. Specifically, the interplay between interlayer bias and the number of cells is identified as a crucial factor influencing gap formation and transmission asymmetry.


                \pacs{72.80.Vp, 73.21.Ac, 73.23.Ad\\
                {\sc Keywords}:  Trilayer graphene, ABC-Stacking, barrier superlattice, energy spectrum, transmissions, Klein tunneling, conductance.}
               
\end{abstract}          

\maketitle

\section{Introduction}
Nearly two decades have passed since the discovery of graphene by A. Geim and K. Novoselov \cite{graphene}, sparking significant interest among physicists. {This has led to a multitude of theoretical \cite{Zhuo} and experimental \cite{Mao} studies being published on this remarkable material}. {This is due to its attractive properties, which make it a promising candidate for numerous electronic \cite{Esteghamat}, optical \cite{Loh} and mechanical \cite{Khan} applications.}. Indeed, besides its two-dimensional structure, graphene boasts high electron mobility \cite{Morozov,Pulizzi,Gosling,Cao}, impressive thermal conductivity \cite{Balandin,Park,Huanga,Calisi}, transparency \cite{Nair,Belyaeva,Nicolosi}, flexibility \cite{Razaq,Shenf} and high mechanical strength \cite{Bhat,Ömer}. {Recently, graphene superlattice structures have shown some negative differential resistance properties \cite{Qijun,Rahighi}}. In addition, within graphene, charge carriers exhibit characteristics similar to those of relativistic massless particles. {They are therefore described by the Dirac equation and have a linear energy dispersion \cite{Kourosh,Ahmed,Gusynin,Katsnelson,Neto,Farjadian}}. Furthermore, graphene is known for its distinctive properties, including an unconventional quantum Hall effect \cite{Gusynin,Stormer}, a minimum conductivity \cite{Katsnelson,Tworzydlo}, a zero band gap energy \cite{Cheung,Peng} and Klein tunneling phenomena \cite{Katsnelson,Koniakhin,Korol,Correa}. However, the last two properties are obstacles to the development of graphene-based electronic devices. Several techniques have been developed to create a gap in graphene. These methods include graphene nanoribbons \cite{Xie,Narita,Aslanidou}, epitaxial substrates \cite{Tringides, Duan}, hydrogenation \cite{Betti,Fei,J. Hong}, strain engineering \cite{Shahid,Zhimei} and biased AB bilayer graphene (BLG) \cite{Van,Hassane,Nadia,Saley}. 

{Spatial non-locality of the optoelectronic response is a common phenomenon observed in many gapped materials. This property is due to the slow recombination of photoexcited carriers. However, a long-range photocurrent response has been observed in gapless systems where carrier recombination occurs rapidly relative to the timescales of carrier diffusion \cite{Song,Mayes,Kiemle}. For example, in graphene, it has been shown that a local current source generates an electric field that scatters surrounding carriers far from the excitation region, explaining the long-range character \cite{Song,Kiemle}.
}

{Recent studies have placed considerable emphasis on trilayer graphene (TLG), with a particular focus on the rhombohedral (ABC) \cite{Guo,Ming,Senthil,Chunli,Shen,Castro,Patri,Qin,Azizimanesh,Tasis} and bernal  (ABA) \cite{Azizimanesh,Tasis,Jia,Ahmadzadeh,Lin} stacking arrangements}. The ABC-TLG displays a cubic energy dispersion characterized by two bands touching at low energy. In contrast, the ABA-TLG stacking configuration features both linear bands akin to monolayer graphene (MLG) and parabolic bands similar to AB-BLG \cite{Lui,Uddin}. Furthermore, it is worth noting that in the presence of a potential barrier, electrons exhibit perfect transmission, i.e., Klein tunneling \cite{Duppen,Ben,El Mouhafid} in {ABC-TLG}, as in MLG. However, as in AB-BLG, the application of an interlayer bias breaks the inversion symmetry in {ABC-TLG}, leading to the emergence of a band gap\cite {Lui,Ben,El Mouhafid,Zou13}. In contrast, the ABA stacking maintains its metallic characteristics under the same constraint. However, the presence of an interlayer bias breaks the mirror symmetry, resulting in the hybridization of the linear and parabolic bands \cite{Ben,El Mouhafid}.

{Note that the possibility of inducing a gap in ABC-TLG makes it a promising material for electronic and optoelectronic devices \cite{Lui,Zou13}. The special properties of ABC-TLG go beyond this. Indeed, it has been shown that ABC-TLG exhibits strong ferromagnetic behavior, which is more robust than in MLG, BLG, and ABA-TLG \cite{Olsen}. In addition, it provides a platform for studying the intricate physics of superconductivity. Indeed, recent studies have reported the observation of superconductivity in ABC-TLG \cite{Watanabe,Awoga}. This may lead to the development of a new generation of field-effect-controlled heterostructures (mesoscopic electronic devices) exploiting correlated electron phenomena \cite{Watanabe}. Furthermore, it was demonstrated in ref. \cite{Lili} that when an ABC-TLG is aligned on a monolayer of hexagonal boron nitride (hBN), the moiré potential induces flat minibands, resulting in Mott insulating behavior at certain band fillings.}

{In a recent study \cite{El Mouhafid} dealing with double barriers in ABC TLG, it was observed that the presence of bound states in the well between the two barriers results in a non-zero transmission inside the gap induced by an interlayer bias. This is not suitable for the development of graphene-based transistors as it does not provide distinct on/off switching states. In this study, we aim to investigate how a periodic structure formed by the alternation of a barrier and a well influences the electron transmission and the gap induced by the bias. Our findings demonstrate that: i) the issue of the appearance of transmission within the gap observed in ref. \cite{El Mouhafid} can be resolved by increasing the number of cells in the superlattice structure, ii) widening the barriers/wells increases the number of gaps, iii) with an increasing number of cells, Klein tunneling decreases while oscillations increase, iv) conductivity decreases with the number of cells and also with increasing barrier height.}

{For further details}, when the energy is less than the interlayer coupling $\gamma_1\approx0.4$ eV, only one propagation mode is available, which results in one channel of transmission. We have shown that increasing the number of cells results in reduced transmission, and oscillations occur at non-normal incidence. In the presence of an interlayer bias, a pseudo-gap is found for a low number of cells ($N=1$) and turns into a gap when the number of cells is increased. It has also been shown that increasing the barrier well/width increases the number of oscillations as well as the number of gaps. For energy greater than $\gamma_1$, three modes are available, resulting in nine transmission channels. In the non-scattered channels, the transmission decreases with the number of cells, the barrier height, and the barrier/well width. In addition, the number of gaps can be increased by increasing the barrier/well width. In the scattered channels, asymmetry with respect to normal incidence ($k_y=0$) is found due to the presence of the interlayer bias. In channels $T^{(1)}_2(k_y)=T^{(2)}_1(-k_y)$ and $T^{(2)}_3(k_y)=T^{(3)}_2(-k_y)$ the transmission becomes more pronounced when the number of cells increases. Moreover, in channels $T^{(1)}_3$ and $T^{(3)}_1$, the transmission is very sensitive to the barrier height. Furthermore, the conductance decreases when the number of cells increases and this results in a gap region. When the barrier height is increased, it has found a displacement in the conductance as well as the gap region.
 

The present paper is outlined as follows: In Sec. \ref{TTMM}, we present a theoretical model and proceed to compute the corresponding eigenvectors and eigenvalues for both two-band and six-band cases. Sec. \ref{Trans} is devoted to determining the transmission probability and conductance based on continuity conditions at the interfaces of the system and the transfer matrix method. In Sec. \ref{RRDD}, we numerically present and discuss our results, comparing them with the existing literature. Finally, in Sec. \ref{CC}, we provide a summary of our findings and draw conclusions from our study.

\begin{figure}[t!]
\vspace{0.cm}
\centering\graphicspath{{./Figures/}}
\includegraphics[width=\linewidth]{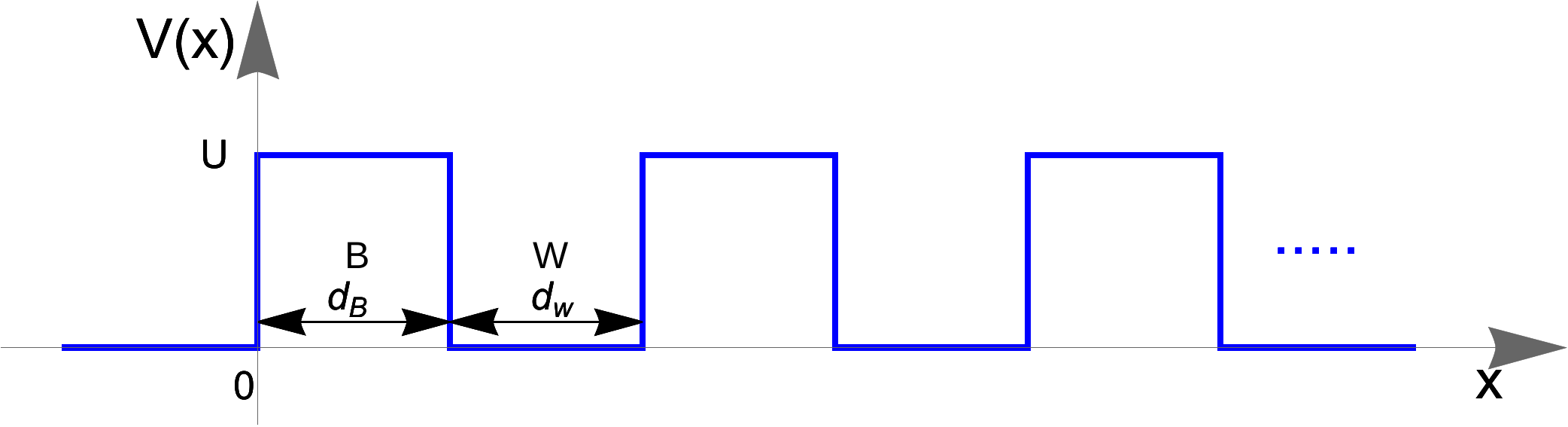}
\caption{(Color online) Schematic diagram of the periodic potential barrier applied to {ABC-TLG}.}\label{Fig1}
\end{figure}
\section{Theory and methods}\label{TTMM}
We consider an {ABC-TLG} subjected to a periodic potential barrier along the $x$-direction {as illustrated in Fig. \ref{Fig1}}. Our system's unit cell comprises two distinct regions: the first is identified as a potential barrier region denoted by $j=B$, and the second is a region with zero potential labeled as $j=W$, referred to as the "well" to distinguish it from the barrier region. The periodic potential is expressed as follows:
\begin{equation}
V_j(x)=\left\{
\begin{array}{ll} 
        U_j\mathbb{I}_s+\Delta_j, & \ \text{for barrier}\\
0, & \ \text{for well} 
\end{array}
\right.
\label{E1}
\end{equation}
{where $U_j$ is the barrier height, $\mathbb{I}_s$ is the $s\times s$ unit matrix (with s=2 and 6 for the two band and six band cases, respectively), $\Delta_j$ stands for the interlayer bias with $\Delta_j=\text{diag}\left(\delta_j,-\delta_j\right)$ for the two-band  and $\Delta_j=\text{diag}\left(\delta_j,\delta_j,0,0,-\delta_j,-\delta_j\right)$ for the six-band, $d_B$ and $d_W$ are, respectively, the barrier and well widths, with the unit cell represented by $d=d_B+d_W$}. 


Initially, we will establish a comprehensive solution for the spinor within the barrier regions. Subsequently, to derive the solution in the well regions, we simply need to enforce \(U_j=\delta_j=0\). To achieve this, we employ the eigenvalue equation \(H_j\Psi_j = E_j\Psi_j\) in conjunction with the Hamiltonians of two-band and six-band cases and the corresponding spinors \(\Psi_j'=\left(\psi_{A_1j},\psi_{B_1j}\right)^\dagger\) and \(\Psi_j=\left(\psi_{A_1j},\psi_{B_1j},\psi_{A_2j},\psi_{B_2j},\psi_{A_3j},\psi_{B_3j}\right)^\dagger\), respectively, with the symbol \(\dagger\) denotes the transpose row vector. Because of translational invariance in both cases, momentum is conserved in the $y$-direction (\([H_j,p_{y}]=0\)), and then we can write 
\begin{align}
&\Psi'_j(x,y)=e^{i k_y y}\left(\psi_{A_1j},\psi_{B_1j}\right)^\dagger
 \label{E2}
\\
&
\Psi_j(x,y)=e^{i k_y y}\left(\psi_{A_1j},\psi_{B_1j},\psi_{A_2j},\psi_{B_2j},\psi_{A_3j},
\psi_{B_3j}\right)^\dagger
\label{E3}.
\end{align}
To simplify, we will adopt the length scale \(a=\frac{\hbar v_F}{\gamma_1}=1.64\) nm along with dimensionless quantities: \(E_j\rightarrow\frac{E_j}{\gamma_1}\), \(U_j\rightarrow \frac{U_j}{\gamma_1}\), \(\delta_j\rightarrow\frac{\delta_j}{\gamma_1}\), \(x\rightarrow\frac{x}{a}\), and \(k_{y}\rightarrow ak_{y}.\)

\subsection{Two-band model}
\label{Twoband1}
{In the regime of very low energy, it becomes feasible to introduce an effective two-band Hamiltonian to describe the ABC-TLG  \cite{Ben, Duppen}. This is} 
\begin{eqnarray}
H'_j&=&\left[
\begin{array}{cc}
U_j+\delta_j  & \upsilon\pi^{\dag3} \\
\upsilon\pi^{3} & U_j-\delta_j %
\end{array}%
\right],\label{ham Two band}
\end{eqnarray}
{where $\upsilon=\frac{\left(\hbar v_{F}\right)^{3}}{\gamma_{1} ^{2}}$, $\pi=q_{x}+ik_{y}$ represents the in-plane momentum, and \(v_{F}=10^{6}\) m/s denotes the Fermi velocity. It is important to note that this approximation holds true only when the Fermi energy is significantly smaller than \(\gamma _{1}\).}
The associated eigenstates solution in each region $j=B$ or $j=W$  can be written in matrix form with plane waves as
\begin{equation}
\Psi'_j(x,y)=\mathcal{L_{\textit{j}}}\mathcal{M_\textit{j}}(x)\Lambda_je^{ik_{y}y},\label{generalsolution}
\end{equation}
and $\mathcal{L_{\textit{j}}}$ is a $6\times2$ matrix  given by
\begin{equation}
\mathcal{L_{\textit{j}}}=  \left[
\begin{array}{cccccccc}
1 & 1 & 1 & 1 & 1  & 1 \\
h_{1j}^{+} & h_{1j}^{-} & h_{2j}^{+}& h_{2j}^{-} & h_{3j}^{+} & h_{3j}^{-}%
\end{array}%
\right],
\end{equation}
{where $h_{\tau j}^{\pm}=\frac{\left(
\pm q_{\tau j}+ik_{y}\right) ^{3}}{\varepsilon_j+\delta_j}$, $\tau=1, 2, 3$ refers to the propagating modes, $\varepsilon_j=E_j-U_j$.  The remaining matrices are 
\begin{align}
&\mathcal{M_\textit{j}}(x)=\text{diag}\left[
e^{i q_{1j}x}, e^{-i q_{1j}x}, e^{i q_{2j}x},\cdots,e^{-i q_{3j}x}\right],
\label{Mjxy}\\
&\Lambda_j=\left[\alpha_{1j}, \beta_{1j}, \alpha_{2j}, \beta_{2j},\alpha_{3j}, \beta_{3j}\right]^{\dagger},
\label{Ajmatrix}
\end{align}
and the wave vectors $q_{\tau j}$ of the plane waves are the solutions of the equation
\begin{equation}
\left( q_{\tau j}^{2}+k_{y}^{2}\right)^{3}+\delta_j^{2}=\varepsilon_j ^{2},
\end{equation}
giving rise to the dispersion relation
\begin{equation}
\varepsilon _{j\pm}=\pm\sqrt{\left( q_{\tau j}^{2}+k_{y}^{2}\right)^{3}+\delta_j^{2}}.
\label{energtwo}
\end{equation}
\vspace{0.5 mm}

\subsection{Six-band model}
\label{Sixband1}
The Hamiltonian near the Dirac point for the six-band  can be written in the tight-binding model as \cite{Ben, El Mouhafid}
\begin{eqnarray}
H_j=\left[
\begin{array}{ccccc}
\mathcal{H} & \Gamma & 0 \\
\Gamma^{\dag } & \mathcal{H} & \Gamma \\
0 & \Gamma^{\dag} & \mathcal{H}%
\end{array}%
\right]+V_j(x)\mathbb{I}_6\label{HamABCfull},
\end{eqnarray}
where the interlayer coupling $\Gamma$ is  given by
\begin{eqnarray}
\Gamma=\left[
\begin{array}{cc}
0 & 0 \\
\gamma_{1} & 0%
\end{array}%
\right]\label{Gamma2},
\end{eqnarray}
and $\mathcal{H}=v_F \vec\sigma\cdot\vec p$ is the Hamiltonian of MLG, with $\vec\sigma=\{\sigma_x,\sigma_y\}$ being the Pauli matrices, $\vec{p}=\left(p_x, p_y\right)$ being the in-plane momentum, and $\mathbb{I}_6$ is an $6\times 6$ unit matrix. When  Eq. (\ref{HamABCfull}) is inserted into the eigenvalue equation alongside the six spinors  Eq. (\ref{E3}), it results in six interrelated differential equations
\begin{subequations}
\begin{align}
-i(\partial_x+k_y)\psi_{B_1j}&=(\varepsilon_j-\delta_j)\psi_{A_1j}
 \label{E4a},\\
     -i(\partial_x-k_y)\psi_{A_1j}&=(\varepsilon_j-\delta_j)\psi_{B_1j}-\psi_{A_2j}
      \label{E4b},\\
   -i(\partial_x+k_y)\psi_{B_2j}&=\varepsilon_j\psi_{A_2j}-\psi_{B_1j}
    \label{E4c},\\
   -i(\partial_x-k_y)\psi_{A_2j}&=\varepsilon_j\psi_{B_2j}-\psi_{A_3j}
    \label{E4d},\\
  -i(\partial_x+k_y)\psi_{B_3j}&=(\varepsilon_j+\delta_j)\psi_{A_3j}-\psi_{B_2j}
   \label{E4e},\\
   -i(\partial_x-k_y)\psi_{A_3j}&=(\varepsilon_j+\delta_j)\psi_{B_3j} \label{E4f}.
\end{align}
\end{subequations}
Eliminating one by one the unknowns in Eqs. (\ref{E4a}-\ref{E4f}), we obtain a six-order differential equation for $\psi_{A_{3}j}$
\begin{widetext}
\begin{equation}
\left[\partial_x^6-\left(\mu_j+3 k_y^2\right) \partial_{x}^4+\left(\nu_j+2 \mu_j k_y^2+3 k_y^4\right)\partial_{x}^{2}-k_y^6-\mu_j k_y^4-\nu_j k_y^2-\rho_j\right] \psi_{A_3j}=0,
\label{E5}
\end{equation}
\end{widetext}
where we have set the parameters 
$\mu_j=-3\varepsilon_j^2-2\delta_j^2, \nu_j=3\varepsilon_j^4+\delta_j^4-2\varepsilon_j^2$, and $\rho_j=-\varepsilon_j^6-\varepsilon_j^2(1+\delta_j^2)^2+2\varepsilon_j^4(\delta_j^2+1)+\delta_j^2$.
The solution of Eq. (\ref{E5}) is a linear combination of plane waves
  \begin{equation}
\psi_{A_3j}=\sum_{\tau=\text{1}}^{\text{3}} \left(\alpha_{\tau j} e^{i k_{\tau j} x}+ \beta_{\tau j} e^{-i k_{\tau j} x} \right),
  \label{E14}
  \end{equation}
where {$k_{\tau j}$} is the waves vector in the x-direction.
By using Eqs.~(\ref{E4a}-\ref{E4f}) and (\ref{E14}), we can derive the remaining components of the spinors and {then express the general solution in matrix form 
\begin{equation}
\Psi_j(x,y)=L_{\textit{j}}M_j(x)\Lambda_j e^{ik_{y}y},\label{six band}
\end{equation}
such as $L_j$ is  given by
\begin{widetext}
\begin{equation}
   L_j=\begin{pmatrix}
               \frac{f_{1j}^{-}(\lambda_{1j} \eta_{1j}-\varepsilon_j b_j)}{c_jb_j f_{1j}^{+}}  &\frac{f_{1j}^{+}(\lambda_{1j} \eta_{1j}-\varepsilon_j b_j)}{c_jb_j f_{1j}^{-}} & \frac{f_{2}^{-}(\lambda_{2j} \eta_{2j}-\varepsilon_j b_j)}{c_jb_j f_{2j}^{+}} &\frac{f_{2j}^{+}(\lambda_{2j} \eta_{2j}-\varepsilon_j b_j)}{c_jb_j f_{2j}^{-}} & \frac{f_{3j}^{-}(\lambda_{3j} \eta_{3j}-\varepsilon_j b_j)}{c_jb_j f_{3j}^{+}} &\frac{f_{3j}^{+}(\lambda_{3j} \eta_{3j}-\varepsilon_j b_j)}{c_jb_j f_{3j}^{-}} \\
               \frac{(\lambda_{1j} \eta_{1,j}-\varepsilon_j b_j)}{b_j f_{1}^{+}}  &-\frac{(\lambda_{1j} \eta_{1j}-\varepsilon_j b_j)}{b_j f_{1j}^{-}} & \frac{(\lambda_{2j} \eta_{2j}-\varepsilon_j b_j)}{b_j f_{2j}^{+}} &-\frac{(\lambda_{2j} \eta_{2j}-\varepsilon_j b_j)}{b_j f_{2j}^{-}} & \frac{(\lambda_{3j} \eta_{3j}-\varepsilon_j b_j)}{b_j f_{3j}^{+}} &-\frac{(\lambda_{3j} \eta_{3j}-\varepsilon_j b_j)}{b_j f_{3j}^{-}} \\
               \frac{(\varepsilon_j\eta_{1,j}-b_j)}{b_j f_{1j}^{+}} &-\frac{(\varepsilon_j \eta_{1j}-b_j)}{b_j f_{1j}^{-}} & \frac{(\varepsilon_j \eta_{2j}-b_j)}{b_j f_{2j}^{+}} &-\frac{(\varepsilon_j \eta_{2j}-b_j)}{b_j f_{2j}^{-}} & \frac{( \varepsilon_j\eta_{3j}-b_j)}{b_j f_{3j}^{+}} &-\frac{(\varepsilon_j\eta_{j}-b_j)}{b_j f_{3j}^{-}}\\
               \frac{\eta_{1j}}{b_j}& \frac{\eta_{1j}}{b_j}& \frac{\eta_{2j}}{b_j}&\frac{\eta_{2j}}{b_j}& \frac{\eta_{3j}}{b_j}&\frac{\eta_{3j}}{b_j}\\
                1 & 1 & 1 & 1&1&1 \\
                 \frac{f_{1j}^{+}}{b_j}& -\frac{f_{1j}^{-}}{b_j}& \frac{f_{2j}^{+}}{b_j}&-\frac{f_{1j}^{-}}{b_j}& \frac{f_{3j}^{+}}{b_j}&-\frac{f_{1j}^{-}}{b_j}&
        \end{pmatrix}
        \label{E16},
\end{equation}
\end{widetext}
and $M_j(x)$ reads as
\begin{equation}
M_j(x)=\text{diag}\left[
e^{i k_{1j}x}, e^{-i k_{1j}x}, e^{i k_{2j}x},\cdots,e^{-i k_{3j}x}\right]
\label{MM}.
\end{equation} 
The matrix $\Lambda_j$ is defined in Eq. (\ref{Ajmatrix}) and  will be specified for the incident and transmitted regions in Sec. \ref{Trans}.
Here we have set $b_j=\varepsilon_j+\delta_j$ and $c_j=\varepsilon_j-\delta_j$ as well as the following parameters
\begin{subequations}
    \begin{align}
        & f_{\tau,j}^{\pm}=k_{\tau j}\pm i k_y
         \label{E17a},\\
        &\eta_{\tau j}=b_j^2-k_{\tau j}^2 -k_y^2
          \label{E17b},\\
        &\lambda_{\tau j}=\varepsilon_j^2-k_{\tau j}^2 -k_y^2
          \label{E17c}.
    \end{align}
\end{subequations}

 Now, to obtain the eigenvalues, we can simply calculate the determinant $\det(H_{j}-\mathbb{I}_6E_j)=0$. This process yields  a polynomial of degree six in terms of $\varepsilon_j$
 \begin{equation}
\varepsilon_j^6+\varepsilon_j^4\kappa_{1j}+\varepsilon_j^2\kappa_{2j}+\kappa_{3j}=0
  \label{E8},
 \end{equation}
 where we have defined the quantities
 \begin{subequations}
  \begin{align}
  & \kappa_{1j}=-3k_j^2-2\delta_j^2-2
  \label{E9a},\\
  &\kappa_{2j}=3k_j^4+2k_j^2+(1+\delta_j^2)^2
  \label{E9b},\\
  &\kappa_{3j}=-\delta_j^2-(k_j^3-k_j\delta_j^2)^2
  \label{E9c},
  \end{align}
 \end{subequations}
and $k_j=\sqrt{k_{\tau j}^2+k_y^2}$ is the wave vector along $x$-direction. To solve Eq.~\eqref{E8}, we apply Cardano's method \cite{Hellesland} to end up with the six-band solutions
 \begin{subequations}\label{E10}
  \begin{align}
  & \varepsilon_{1j}^{\pm}= \pm \sqrt{2\sqrt{\Xi_j}\cos{\left[\frac{\vartheta_j+2\pi}{3}\right]}-\frac{\kappa_{1j}}{3}}
  \label{E10a},\\
  & \varepsilon_{2j}^{\pm}= \pm \sqrt{2\sqrt{\Xi_j}\cos\left[\frac{\vartheta_j+4\pi}{3}\right]-\frac{\kappa_{1j}}{3}}
  \label{E10b},\\
  & \varepsilon_{3j}^{\pm}=  \pm \sqrt{2\sqrt{\Xi_j}\cos\left[\frac{\vartheta_j}{3}\right]-\frac{\kappa_{1j}}{3}}
  \label{E10c},
  \end{align}
 \end{subequations}
 with the set of the following parameters $\Xi_j=\frac{\kappa_{1j}^2-3\kappa_{2j}}{9}$, $\vartheta_j=\arccos\left(\Theta_j\sqrt{\frac{1}{\Xi_j^3}}\right)$ and $\Theta_j=\frac{9\kappa_{1j}\kappa_{2j}-27\kappa_{3j}-2\kappa_{1j}^3}{54}$.

By using the six bands Eq. (\ref{E10}), we can derive six longitudinal waves vectors { $k_{\tau j}$} $(\tau=1,2,3)$ for propagating modes. They are given by
 \begin{subequations}
  \begin{align}
  & k_{1j}^{\pm}= \pm \sqrt{2\sqrt{\omega_j}\cos\left[\frac{\varpi_j+2\pi}{3}\right]-\frac{\mu_{j}}{3}-k_y^2} \label{E12a},\\
  & k_{2j}^{\pm}= \pm \sqrt{2\sqrt{\omega_j}\cos\left[\frac{\varpi_j+4\pi}{3}\right]-\frac{\mu_{j}}{3}-k_y^2}\label{E12b},\\
  &k_{3j}^{\pm}=  \pm \sqrt{2\sqrt{\omega_j}\cos\left[\frac{\varpi_j}{3}\right]-\frac{\mu_{j}}{3}-k_y^2}\label{E12c},
  \end{align}
 \end{subequations}
 where we have defined the quantities $\omega_j=\frac{\mu_{j}^2-3\nu_{j}}{9}$, $\varpi_j=\arccos\left(\varrho_j\sqrt{\frac{1}{\omega_j^3}}\right)$ and $\varrho_j=\frac{9\mu_{j}\nu_{j}-27\rho_{j}-2\mu_{j}^3}{54}$.
  
\begin{figure}[t!]
\vspace{0.cm}
\centering\graphicspath{{./Figures/}}
\includegraphics[width=\linewidth]{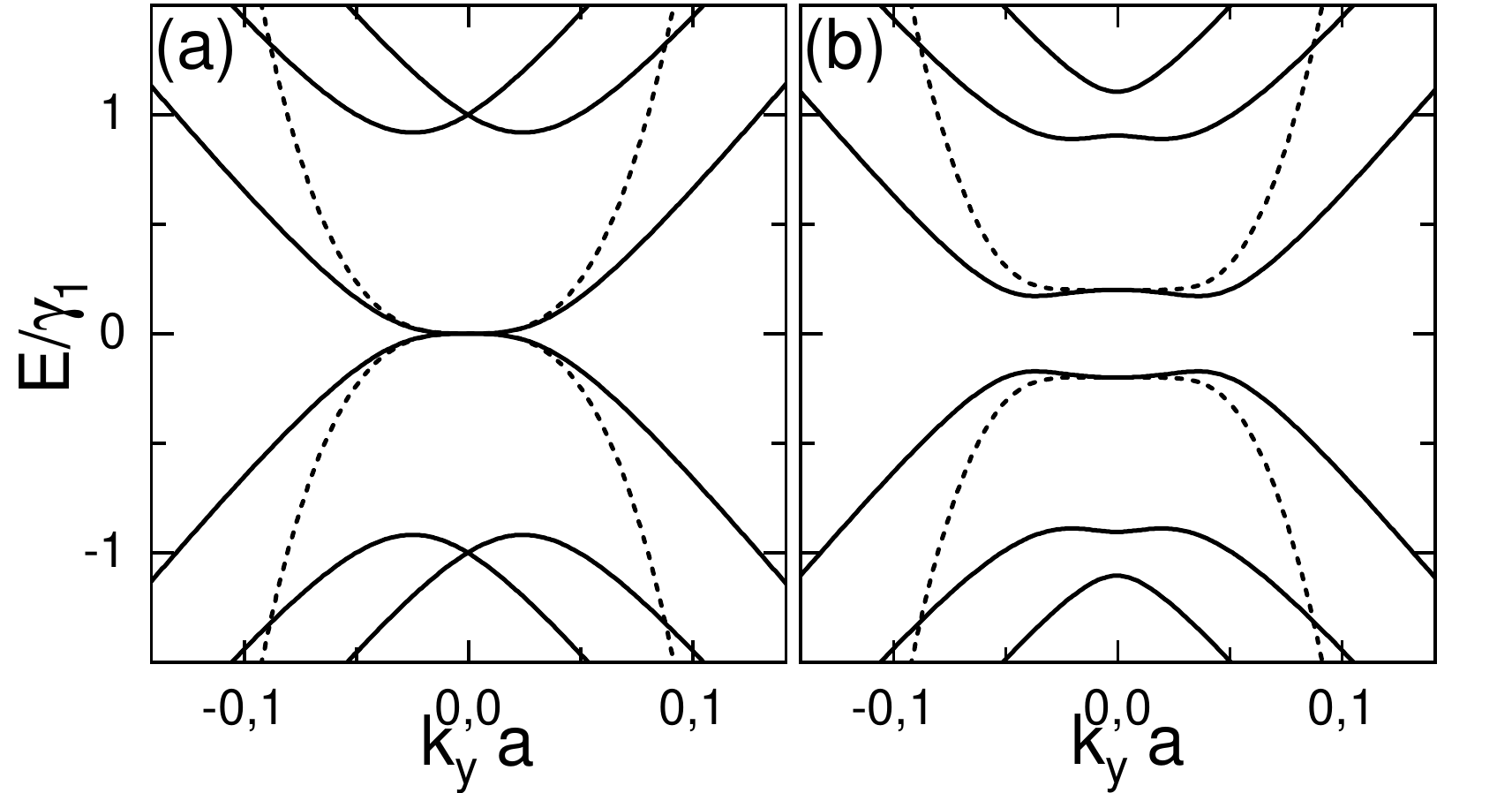}
\caption{(Color online) Energy spectrum {of ABC-TLG} as a function of the  transverse wave vector $k_y$. The dashed lines correspond to the spectrum of the two-band Hamiltonian, while the solid lines correspond to the six-band Hamiltonian. We choose (a): $\delta=0$, and (b): $\delta=0.2\gamma_1$.}\label{energy}
\end{figure}

\subsection{Discussions}
\label{Discussion}

In Fig. \ref{energy} we show the dispersion relation obtained in the two band model (dashed curves ) and the dispersion relation obtained from the six band model (solid curves). We clearly see that for $E\ll \gamma_1$ and $\delta=0$, we have two bands  that touch each other at $E=0$, as shown in Fig. \ref{energy}(a). For energy exceeding $\gamma_1$, we have six bands, two of which touch at $E=0$. The four others are separated by a gap $\gamma_1$ from those that touch at $E=0$ \cite{Rondeau}. In the presence of an interlayer bias $\delta\neq 0$ in Fig. \ref{energy}(b), the two touched bands at $E=0$ are separated, and a gap is opened between them in both the two-band and six-band cases. In addition, the bands that cross at high energy in the six-band case are also separated. {The band structures of the ABC-TLG superlattice have been studied in \cite{Uddin,Salah}, and therefore, we will focus on the tunneling effect in the following.
}
 
\section{Transmission and conductance} \label{Trans}
We will apply the spinor continuity conditions at each interface of the system and employ the transfer matrix method to calculate the transmission probability. Initially, we will define the wave function $\Psi_j$ in the incident and transmitted regions, ensuring equivalence through the relation $L_{\text{in}} M_{\text{in}}(x)=L_{\text{tr}} M_{\text{tr}}(x)$ and $k_{\tau,\text{in}}=k_{\text{tr}}$ due to zero potential and interlayer bias. Consequently, we can eliminate the subscripts from matrices $M(x)$ and $L$ to express the wave functions as
\begin{equation}
    \Psi_{\text{in}/\text{tr}}=L  M(x) \Lambda_{\text{in}/\text{tr}}
    \label{Eqq20},
\end{equation}
where $\Lambda_{\text{in}}$ in the incidence and $\Lambda_{\text{tr}}$ in the transmission are provided for the six band case by
\begin{align}
&       \Lambda_{\text{in}}= \left(\delta_{\tau 1},
        r_{1}^{\tau},
        \delta_{\tau 2},
        r_{2}^{\tau},
        \delta_{\tau 3},
       r_{3}^{\tau}\right)^\dagger,\\
& 
\Lambda_{\text{tr}}=
\left(t_{1}^{\tau}, 0,
                t_{2}^{\tau},
                0,
                 t_{3}^{\tau},
                0\right)^\dagger
                \label{Eq32},
                \end{align}
 where $\delta_{\tau,i}$ (where $i=1,2,3$) represents the Kronecker symbol, signifying the right-propagating or evanescent waves. $r_{i}^{\tau}$ corresponds to the left-propagating reflected waves, while $t_{i}^{\tau}$ pertains to the right-propagating transmitted waves. {By matching the wave functions $\Psi_j$ at the interface of each region, the coefficients in the incidence region $\Lambda_{\text{in}}$ and those in the transmission region $\Lambda_{\text{tr}}$ can be related through the transfer matrix $\chi$ in the following manner}
\begin{equation}
   \Lambda_{\text{in}}=\chi \Lambda_{\text{tr}}
    \label{E20},
\end{equation}
and $\chi$ is given by
 \begin{equation}
\chi= M^{-1}(0) L^{-1} \Omega^{n} L M(nd)
\label{EN},
\end{equation}
where  $n$ is the number of cells and $\Omega$ is 
\begin{equation}
    \Omega=L_B M_B(0)M_B^{-1}(d_B)L_B^{-1}L_W M_W(d_B)M_W^{-1}(d)L_W^{-1}.
\end{equation}
Then, after some algebras, we can write Eq. (\ref{E20}) in the form
\begin{widetext}
\begin{equation}
\begin{pmatrix}
        t_{1}^{\tau} \\
        r_{1}^{\tau} \\
        t_{2}^{\tau} \\
        r_{2}^{\tau} \\
        t_{3}^{\tau} \\
        r_{3}^{\tau}
\end{pmatrix}=\begin{pmatrix}
        \chi_{11} & 0 & \chi_{13} & 0& \chi_{15}&0\\
        \chi_{21} & -1 & \chi_{23} & 0 &\chi_{25}&0  \\
        \chi_{31} & 0 & \chi_{33} & 0& \chi_{35} & 0\\
        \chi_{41} & 0 & \chi_{43} & -1 &\chi_{45}& 0\\
        \chi_{51} & 0 & \chi_{53} & 0& \chi_{55} & 0\\
        \chi_{61} & 0 & \chi_{63} & 0 &\chi_{65}& -1\\
\end{pmatrix}^{-1} \begin{pmatrix}
        \delta_{\tau, 1} \\
        0 \\
        \delta_{\tau,2} \\
        0\\
         \delta_{\tau,3} \\
        0
\end{pmatrix}
\label{23},
\end{equation}
\end{widetext}  
where $\chi_{uv}$ are elements of matrix $\chi$. Thereafter, we can derive the expression of the transmission coefficients through Eq. (\ref{23}) to get the followings
\begin{widetext}
\begin{align}
&t_{1}^{\tau}=\frac{\left(\chi_{35}\chi_{53} -\chi_{33}\chi_{55}\right)\delta_{\tau, 1}+\left(\chi_{13}\chi_{55}-\chi_{15}\chi_{53}\right)\delta_{\tau, 2}+\left(\chi_{15}\chi_{33} -\chi_{13}\chi_{35}\right)\delta_{\tau,3}}{\chi_{15} \chi_{33} \chi_{51}-\chi_{13} \chi_{35} \chi_{51}-\chi_{15} \chi_{31} \chi_{53}+\chi_{11} \chi_{35} \chi_{53}+\chi_{13} \chi_{31} \chi_{55}-\chi_{11} \chi_{33} \chi_{55}},\\
&t_{2}^{\tau}=\frac{\left(\chi_{31}\chi_{55}-\chi_{35}\chi_{51}\right)\delta_{\tau,1}+\left(\chi_{15}\chi_{51}-\chi_{11}\chi_{55}\right)\delta_{\tau,2}+\left(\chi_{11}\chi_{35}-\chi_{15}\chi_{31}\right)\delta_{\tau,3}}{\chi_{15} \chi_{33} \chi_{51}-\chi_{13} \chi_{35} \chi_{51}-\chi_{15} \chi_{31} \chi_{53}+\chi_{11} \chi_{35} \chi_{53}+\chi_{13} \chi_{31} \chi_{55}-\chi_{11} \chi_{33} \chi_{55}},\\
&t_{3}^{\tau}=\frac{\left(\chi_{33}\chi_{51}-\chi_{31}\chi_{53}\right)\delta_{\tau,1}+\left(\chi_{11}\chi_{53}-\chi_{13}\chi_{53}\right)\delta_{\tau,2}+\left(\chi_{13}\chi_{31} -\chi_{11}\chi_{33}\right)\delta_{\tau,3}}{\chi_{15} \chi_{33} \chi_{51}-\chi_{13} \chi_{35} \chi_{51}-\chi_{15} \chi_{31} \chi_{53}+\chi_{11} \chi_{35} \chi_{53}+\chi_{13} \chi_{31} \chi_{55}-\chi_{11} \chi_{33} \chi_{55}}.
\end{align}
\end{widetext}
The process is identical for the two-band case, albeit with the additional consideration of the continuity of the first and second derivatives of the wave functions at the interfaces \cite{Van,Duppen}.
At this point, we need to utilize the current density to calculate the transmission probabilities. It is given by
\begin{equation}
    J=v_F \Psi ^{\dagger} \alpha \Psi
    \label{34},
\end{equation}
where $\alpha$ is a $6\times 6$ matrix with Pauli matrix $\sigma_x$ on its diagonal. We are particularly interested in the current density within the incident and transmitted regions. Thus we inject Eq. (\ref{Eqq20}) into Eq. (\ref{34}) to get
\begin{equation}
    J_{\text{in/tr}}=v_F \Lambda_{\text{in/tr}}^{\dagger}M(x)^{\dagger}A M(x) \Lambda_\text{in/tr}
\end{equation} 
where $A=L^{\dagger}\alpha L$ is a diagonal matrix consisting of traceless \(2 \times 2\) blocks, each one corresponds to a propagation mode. It's important to highlight that since we are not calculating reflection probabilities, $J_\text{in}$ exclusively considers the right-propagating waves $\delta_{\tau,i}$ for $\Lambda_\text{in}$. Consequently, our transmission probabilities are expressed as follows: 
\begin{equation}
  T^{\tau}_{i}=\frac{\left|{j}_{\text{tr}}\right|}{\left|{j}_{\text{in}}\right|}=\frac{A_{ii}}{A_{\tau \tau }}\left|t_{i}^{\tau}\right|^{2}
\label{35},
\end{equation}  
where $A_{ii}$ and $A_{\tau \tau}$ are elements of matrix $A$. It is worth noting that in the two-band case, there exists a single transmission channel from state $\tau$ to state $i$ with indices $\tau=i=1$, and the corresponding transmission probability is denoted as $T\neq0$ \cite{Ben,Duppen,El Mouhafid}.
Now, by using Landauer Butikker's formula \cite{Landauer} we can calculate the conductance in terms of transmission. Thus, we will see how it will be affected by the number of cells. It is given by 
\begin{equation}\label{eq24}
G(E)=G_{0}\frac{L_y}{2
\pi}\int_{-\infty}^{+\infty}dk_{y}\sum^3_{\tau,i=1}T^{\tau}_{i}(E,k_y),
\end{equation}
where $G_0=4e^2/h$ (the factor 4 emerges from graphene's valley and spin degeneracy), and $L_y$ refers to the length of the sample in the y-direction. 

\section{Numerical results}\label{RRDD}
{The analytical results obtained previously have been used in Mathematica to generate numerical results, which we will now elaborate on in this section.}
Two cases will be presented, depending on the energy value considered. When energy is less than $\gamma_1$, only one propagation mode is available, corresponding to the two-band case. 
When energy exceeds $\gamma_1$, three propagation modes are available, leading to nine transmission channels. Given that the potential and interlayer bias are zero in the well regions ($U_j=\delta_j=0$ when $j=W$), we will simplify notation by omitting the subscript $j$. Instead, we will use the symbols $U$ and $\delta$ to denote the potential and interlayer bias, respectively, in the barrier regions throughout the discussion. 
\subsection{Two band tunneling}

In Fig. \ref{Fig3}, we plot the transmission as a function of energy $E$ and the transverse wave vector $k_y$ for barrier height $U=0.5\gamma_1$ and barrier/well width of $d_B=d_W=2$nm. Fig. \ref{Fig3}(a) illustrates the transmission for $N=1$ in the absence of an interlayer bias ($\delta=0$). As found in refs. \cite{Duppen,Ben,El Mouhafid}, Klein tunneling occurs at normal incidence due to the conservation of pseudospin in {ABC-TLG}. In addition, for energy $E<0.4\gamma_1$, Klein tunneling is observed at non-normal incidence, which is in contrast with the findings in refs. \cite{Ben,El Mouhafid} where the transmission is zero. This results from the low barrier width that we have considered in contrast to refs. \cite{Ben,El Mouhafid}, so in our case, there is a non-zero probability for the evanescent wave to get through the barrier, as indicated in ref. \cite{Xu}. In Fig. \ref{Fig3}(b), we plot the transmission for $N=8$ and show that as the number of cells increases, Klein tunneling narrows at normal incidence and disappears completely when moving away from normal incidence, as the result in refs. \cite{Ben,El Mouhafid}. However, at certain energies, narrowed resonances are observed at nearly normal incidence, as stated in the single barrier case \cite{Ben} and the double symmetric barrier case \cite{El Mouhafid}. On the other hand, the transmission shows oscillations greater than those observed in single and double barrier cases \cite{Ben,El Mouhafid}. Indeed, as the number of cells increases, the number of oscillations also increases, as demonstrated in Fig. \ref{Fig3}(c) for $N=16$. Similar observations have been reported in refs. \cite{Xu,Sánchez} dealing on MLG superlattices. However, it should be noted that these oscillations do not occur at normal incidence, so Klein tunneling remains intact, as shown in refs. \cite{Xu,Sánchez,Cervantes}. To investigate the effect of an interlayer bias, we plot the transmission in Fig. \ref{Fig3}(d) with the same conditions as in Fig. \ref{Fig3}(a), but now we have set $\delta=0.15\gamma_1$. As it can be seen, the presence of $\delta$ results in a reduction of Klein tunneling, both at normal and non-normal incidence, such that a pseudo-gap is formed at energy $E=0.1\gamma_1$. However, a true gap isn't achieved due to the narrow barrier width and the lowest number of cells ($N=1$) \cite{Xu}. When the number of cells is increased, a gap is opened in Fig. \ref{Fig3}(e) and Fig. \ref{Fig3}(f) for $N=8$ and $N=16$, respectively, which is consistent with the results in ref. \cite{Xu}. It is important to note that the evidence of gap induced by a non zero bias in ABC-TLG is observed experimentally in ref \.  In addition, due to the presence of the interlayer bias $\delta$, oscillations appear even at normal incidence, reducing Klein tunneling.
In Fig. \ref{Fig3}, we plot the transmission as a function of energy $E$ and the transverse wave vector $k_y$ for barrier height $U=0.5\gamma_1$ and barrier/well width of $d_B=d_W=2$nm. Fig. \ref{Fig3}(a) illustrates the transmission for $N=1$ in the absence of an interlayer bias ($\delta=0$). As found in refs. \cite{Duppen,Ben,El Mouhafid}, Klein tunneling occurs at normal incidence due to the conservation of pseudospin in {ABC-TLG}. In addition, for energy $E<0.4\gamma_1$, Klein tunneling is observed at non-normal incidence, which is in contrast with the findings in refs. \cite{Ben,El Mouhafid} where the transmission is zero. This results from the low barrier width that we have considered in contrast to refs. \cite{Ben,El Mouhafid}, so in our case, there is a non-zero probability for the evanescent wave to get through the barrier, as indicated in ref. \cite{Xu}. In Fig. \ref{Fig3}(b), we plot the transmission for $N=8$ and show that as the number of cells increases, Klein tunneling narrows at normal incidence and disappears completely when moving away from normal incidence, as the result in refs. \cite{Ben,El Mouhafid}. However, at certain energies, narrowed resonances are observed at nearly normal incidence, as stated in the single barrier case \cite{Ben} and the double symmetric barrier case \cite{El Mouhafid}. On the other hand, the transmission shows oscillations greater than those observed in single and double barrier cases \cite{Ben,El Mouhafid}. Indeed, as the number of cells increases, the number of oscillations also increases, as demonstrated in Fig. \ref{Fig3}(c) for $N=16$. Similar observations have been reported in refs. \cite{Xu,Sánchez} dealing on MLG superlattices. However, it should be noted that these oscillations do not occur at normal incidence, so Klein tunneling remains intact, as shown in refs. \cite{Xu,Sánchez,Cervantes}. To investigate the effect of an interlayer bias, we plot the transmission in Fig. \ref{Fig3}(d) with the same conditions as in Fig. \ref{Fig3}(a), but now we have set $\delta=0.15\gamma_1$. As it can be seen, the presence of $\delta$ results in a reduction of Klein tunneling, both at normal and non-normal incidence, such that a pseudo-gap is formed at energy $E=0.1\gamma_1$. However, a true gap isn't achieved due to the narrow barrier width and the lowest number of cells ($N=1$) \cite{Xu}. When the number of cells is increased, a gap is opened in Fig. \ref{Fig3}(e) and Fig. \ref{Fig3}(f) for $N=8$ and $N=16$, respectively, which is consistent with the results in ref. \cite{Xu}. It is important to note that the evidence of gap induced by a non zero bias in ABC-TLG is observed experimentally in ref \.  In addition, due to the presence of the interlayer bias $\delta$, oscillations appear even at normal incidence, reducing Klein tunneling.
In Fig. \ref{Fig3}, we plot the transmission as a function of energy $E$ and the transverse wave vector $k_y$ for barrier height $U=0.5\gamma_1$ and barrier/well width of $d_B=d_W=2$nm. Fig. \ref{Fig3}(a) illustrates the transmission for $N=1$ in the absence of an interlayer bias ($\delta=0$). As found in refs. \cite{Duppen,Ben,El Mouhafid}, Klein tunneling occurs at normal incidence due to the conservation of pseudospin in {ABC-TLG}. In addition, for energy $E<0.4\gamma_1$, Klein tunneling is observed at non-normal incidence, which is in contrast with the findings in refs. \cite{Ben,El Mouhafid} where the transmission is zero. This results from the low barrier width that we have considered in contrast to refs. \cite{Ben,El Mouhafid}, so in our case, there is a non-zero probability for the evanescent wave to get through the barrier, as indicated in ref. \cite{Xu}. In Fig. \ref{Fig3}(b), we plot the transmission for $N=8$ and show that as the number of cells increases, Klein tunneling narrows at normal incidence and disappears completely when moving away from normal incidence, as the result in refs. \cite{Ben,El Mouhafid}. However, at certain energies, narrowed resonances are observed at nearly normal incidence, as stated in the single barrier case \cite{Ben} and the double symmetric barrier case \cite{El Mouhafid}. On the other hand, the transmission shows oscillations greater than those observed in single and double barrier cases \cite{Ben,El Mouhafid}. Indeed, as the number of cells increases, the number of oscillations also increases, as demonstrated in Fig. \ref{Fig3}(c) for $N=16$. Similar observations have been reported in refs. \cite{Xu,Sánchez} dealing on MLG superlattices. However, it should be noted that these oscillations do not occur at normal incidence, so Klein tunneling remains intact, as shown in refs. \cite{Xu,Sánchez,Cervantes}. To investigate the effect of an interlayer bias, we plot the transmission in Fig. \ref{Fig3}(d) with the same conditions as in Fig. \ref{Fig3}(a), but now we have set $\delta=0.15\gamma_1$. As it can be seen, the presence of $\delta$ results in a reduction of Klein tunneling, both at normal and non-normal incidence, such that a pseudo-gap is formed at energy $E=0.1\gamma_1$. However, a true gap isn't achieved due to the narrow barrier width and the lowest number of cells ($N=1$) \cite{Xu}. When the number of cells is increased, a gap is opened in Fig. \ref{Fig3}(e) and Fig. \ref{Fig3}(f) for $N=8$ and $N=16$, respectively, which is consistent with the results in ref. \cite{Xu}. {It is crucial to recognize that experimental evidence for a gap induced by a non-zero bias in ABC-TLG has been documented in \cite{Lui,Zou13}}. In addition, due to the presence of the interlayer bias $\delta$, oscillations appear even at normal incidence, reducing Klein tunneling.

\begin{figure}[tbh]
\begin{center}
\end{center}
\includegraphics[width=\linewidth]{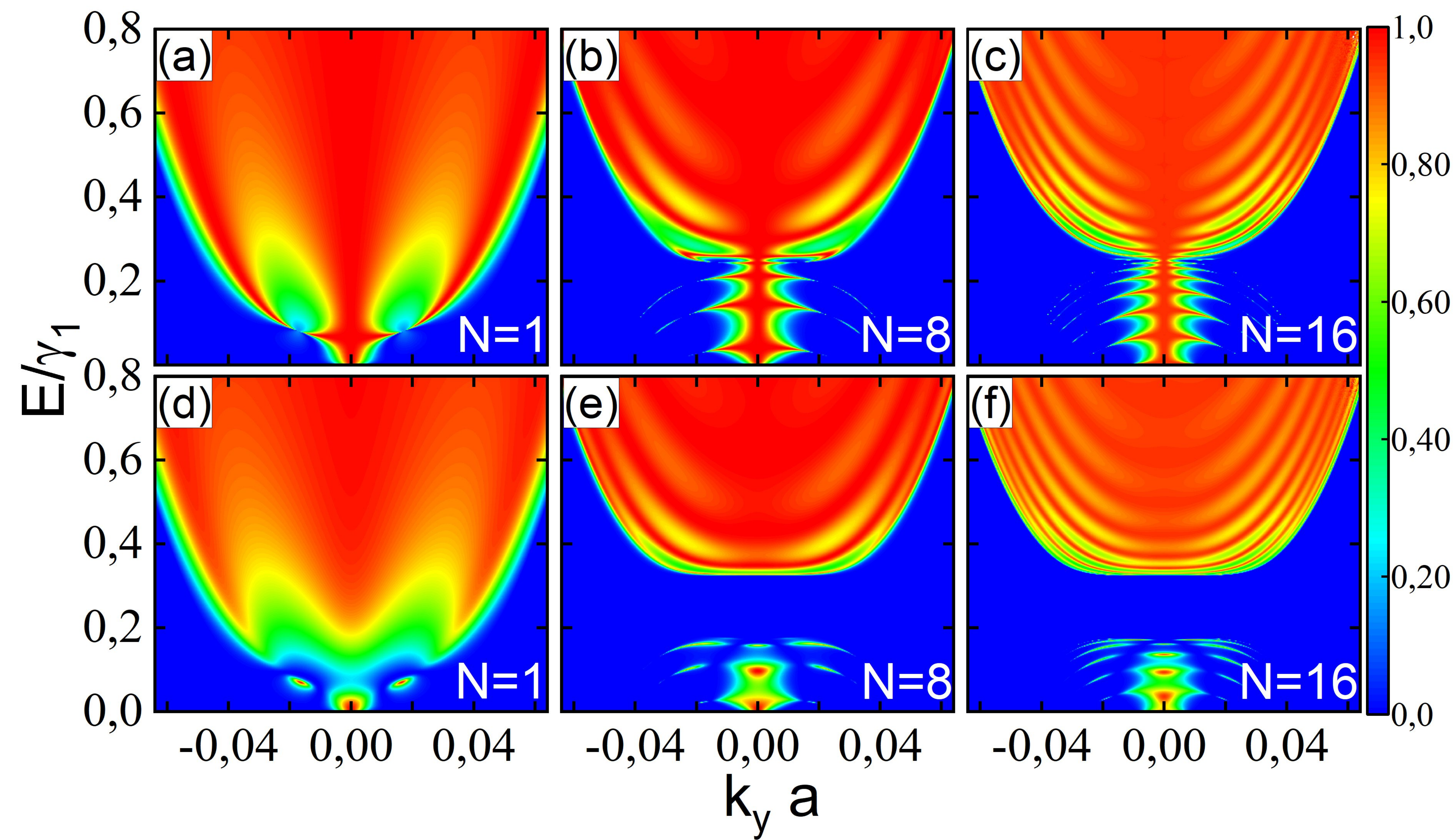}
\caption{(Color online) Transmission {$T$ of ABC-TLG} as a function of energy $E$ and the transverse wave vector $k_y$ for $U=0.5\gamma_1$ and $d_B=d_W=2$ nm. In (a), (b), and (c), we have set the interlayer bias at $\delta=0$, while for (c), (d), and (e), it is set at $\delta=0.15\gamma_1$.}
\label{Fig3}
\end{figure}

\begin{figure}[ht]
      \centering
\includegraphics[width=\linewidth]{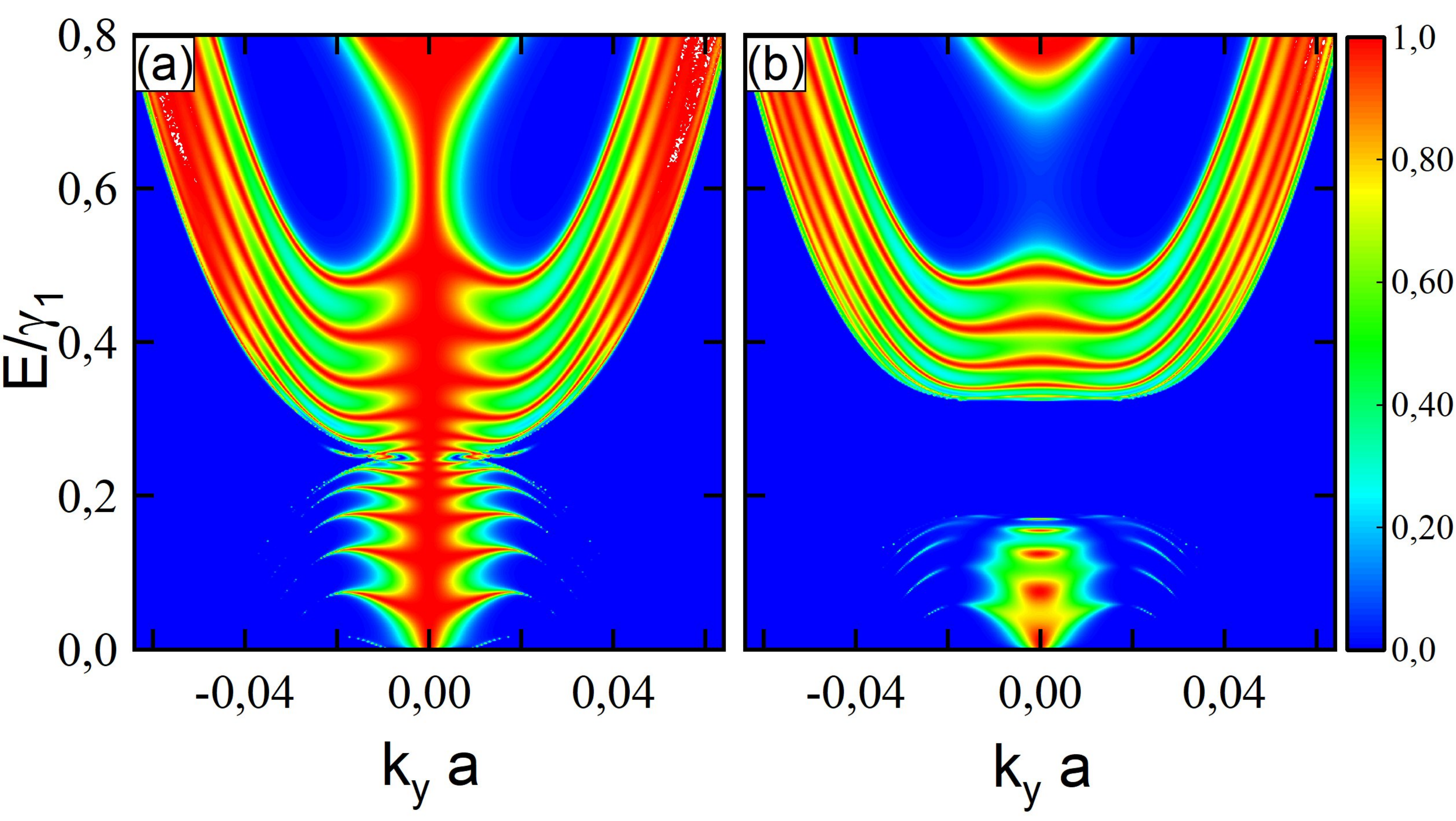}
     \caption{(Color online) The same parameters as in Figs. \ref{Fig3}(b) and (e), respectively, in (a) and (b), but now for $d_B=d_W=4$nm. }
       \label{Fig4}
\end{figure}
In Figs. \ref{Fig4}(a) and (b), we plot the transmission by using the same parameters as in Figs. \ref{Fig3}(b) and (e), respectively, but for increasing barrier/well width, $d_B=d_W=4$nm. Fig. \ref{Fig4}(a) shows the transmission for zero interlayer bias, and we notice that increasing the barrier/well width has a significant influence on the transmission behavior. Indeed, the transmission decreases strongly at non-normal incidence, thus increasing anti-Klein tunneling (zero transmission). In addition, a large number of oscillations are found at non-normal incidence. These oscillations are greater than those obtained in Fig. \ref{Fig3}(b) for $d_B=d_W=2$nm. However, as observed in Figs. \ref{Fig3}(b,c) as well as in refs. \cite{Xu,Sánchez,Cervantes,Alvarado}, these oscillations do not occur at normal incidence. In Fig. \ref{Fig4}(b), when an interlayer bias of $\delta=0.15\gamma_1$ is applied, two gaps are observed in contrast with the results presented in Figs. \ref{Fig3}(e,f) for small barrier/well width. We notice that one of the gaps is wider than the other, and the transmission is shaped like a "Mexican hat" between them. Furthermore, it is noteworthy that Klein tunneling is diminished at normal incidence and oscillations occur, as in Figs. \ref{Fig3}(e,f).

\subsection{Six-band tunneling}
As mentioned above, once the energy exceeds $\gamma_1$, three propagation modes ($k_1$, $k_2$, and $k_3$) become accessible. This, in turn, results in a total of nine transmission channels. Among these channels, three of them, $T^{(1)}_{1}$, $T^{(2)}_{2}$, and $T^{(3)}_{3}$, represent the non-scattered transmission channels, corresponding to propagation via $k_1$, $k_2$, and $k_3$, respectively. The remaining channels, denoted as $T^\tau_i$ with $\tau\neq i$, represent the scattered transmission channels where electrons incoming through one mode are transmitted via another one. We have already seen in the two-band case that, in the absence of an interlayer bias $\delta$, Klein tunneling persists at normal incidence. So in the six-band case, we plot the transmission by varying the other parameters and setting the interlayer bias at $\delta=0.2\gamma_1$.

In Fig.~\ref{DensityPlotN=1.pdf}, we present a density plot of the nine transmission channels as a function of energy $E$ and the transverse wave vector $k_y$. The parameters include a barrier height of $U=2\gamma_1$, a barrier/well width of $d_B=d_W=0.2$nm and a number of cells, $N=1$. In the non-scattered channels $T_{1}^{(1)}$, $T_{2}^{(2)}$ and $T_{3}^{(3)}$ we have full transmission at both normal and non-normal incidence. At normal incidence, this is the manifestation of Klein tunneling and does not depend on the barrier width \cite{Ben}. At non-normal incidence, the full transmission is related to the barrier width. As discussed in ref. \cite{Xu}, the evanescent waves still have a certain probability of being transmitted through the barrier when its width is small. Recall that the condition for having propagation modes $k_2$ and $k_3$ is that the energy exceeds $\gamma_1$, as a result, the transmission occurs only for $E>\gamma_1$ in $T_{2}^{(2)}$ and $T_{3}^{(3)}$ channels. On the other hand, there is no scattering between the three modes $k_1$, $k_2$, and $k_3$, which results in zero transmission in all the scattered channels.


\begin{figure}[tbh]
\begin{center}
\end{center}
\includegraphics[width=\linewidth]{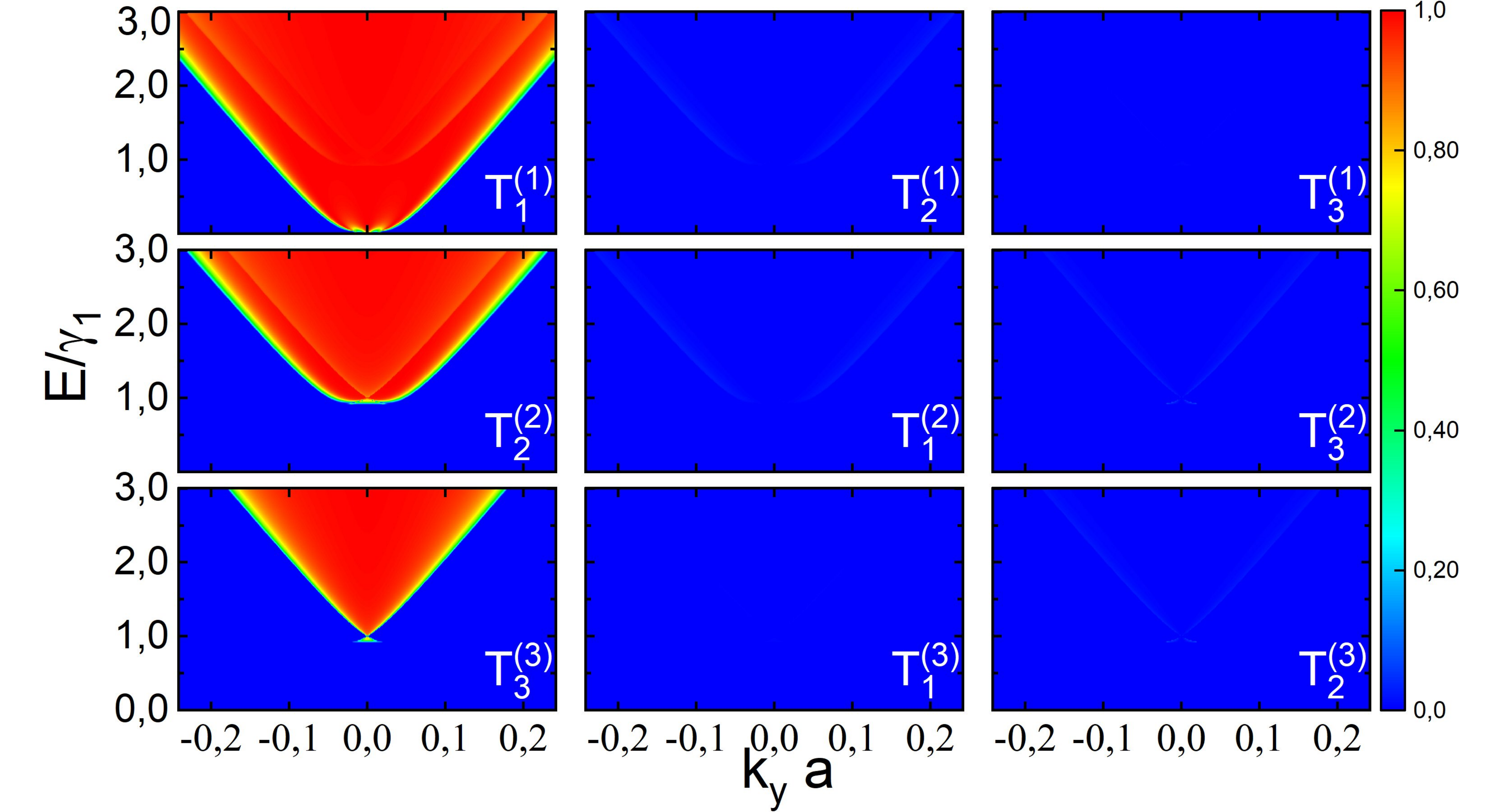}
\caption{(Color online)  Transmission {of ABC-TLG} as a function of incident  energy $E$ and the transverse wave vector $k_y$ for barrier height $U=2\gamma_1$, $d_B=d_W=0.2$nm, and $\delta=0.2\gamma_1$. The number of cells is set at $N=1$. }
\label{DensityPlotN=1.pdf}
\end{figure}


Fig. \ref{DensityPlotN=5.pdf} shows the transmission with the same parameters as in Fig. \ref{DensityPlotN=1.pdf} except that we increase the number of cells at $N=5$. Within the channel $T_{1}^{(1)}$, we notice that the transmission decreases and anti-Klein tunneling occurs at non-normal incidence, in contrast to the result in Fig. \ref{DensityPlotN=1.pdf}. In addition, due to the presence of the interlayer bias $\delta=0.2\gamma_1$, we observe a cut-off in Klein tunneling at normal incidence (transmission decreases from unit to half of unit), which contrasts with the result in ref. \cite{Ben}, where $\delta=0$. However, a gap is not obtained due to the low number of cells and small barrier width, similar to the result in ref. \cite{Xu}. In both channels $T^{(2)}_{2}$ and $T_{3}^{(3)}$, the transmission decreases considerably when we compare with the results in Fig. \ref{DensityPlotN=1.pdf} for $N=1$. As a result, at normal incidence, Klein tunneling starts to appear from $E=U$, where, as in Fig. \ref{DensityPlotN=1.pdf}, it occurs from $E=U-\gamma_1$. In contrast to the results in Fig. \ref{DensityPlotN=1.pdf}, there is a scattering between the modes $k_1$ and $k_2$ ($k_2$ and $k_3$), which results in a weak transmission in the channels $T^{(1)}_{2}$ and $T^{(2)}_{1}$ ( $T^{(2)}_{3}$ and $T^{(3)}_{2}$). However, there is no scattering between $k_1$ and $k_3$, and the transmission is still zero in the channels $T^{(1)}_{3}$ and $T^{(3)}_{1}$. In addition, the presence of the interlayer bias results in an asymmetry with respect to the incidence normal $k_y=0$ in the scattered channels, as found in refs. \cite{El Mouhafid,Van,Nadia}. Moreover, the scattered channels are equal under the change of the sign of $k_y$, such that we have $T^{(1)}_{2}(k_y)=T^{(2)}_{1}(-k_y)$ and $T^{(2)}_{3}(k_y)=T^{(3)}_{2}(-k_y)$. This is a consequence of the time reversal symmetry of the system \cite{Ben}. For example, an electron that is scattered from $k_1$ to $k_2$, near the Dirac point $K$, is equivalent to another that is scattered from $k_2$ to $k_1$, near the Dirac point $K'$ \cite{Ben,Van}.

\begin{figure}[tbh]
\begin{center}
\end{center}
\includegraphics[width=\linewidth]{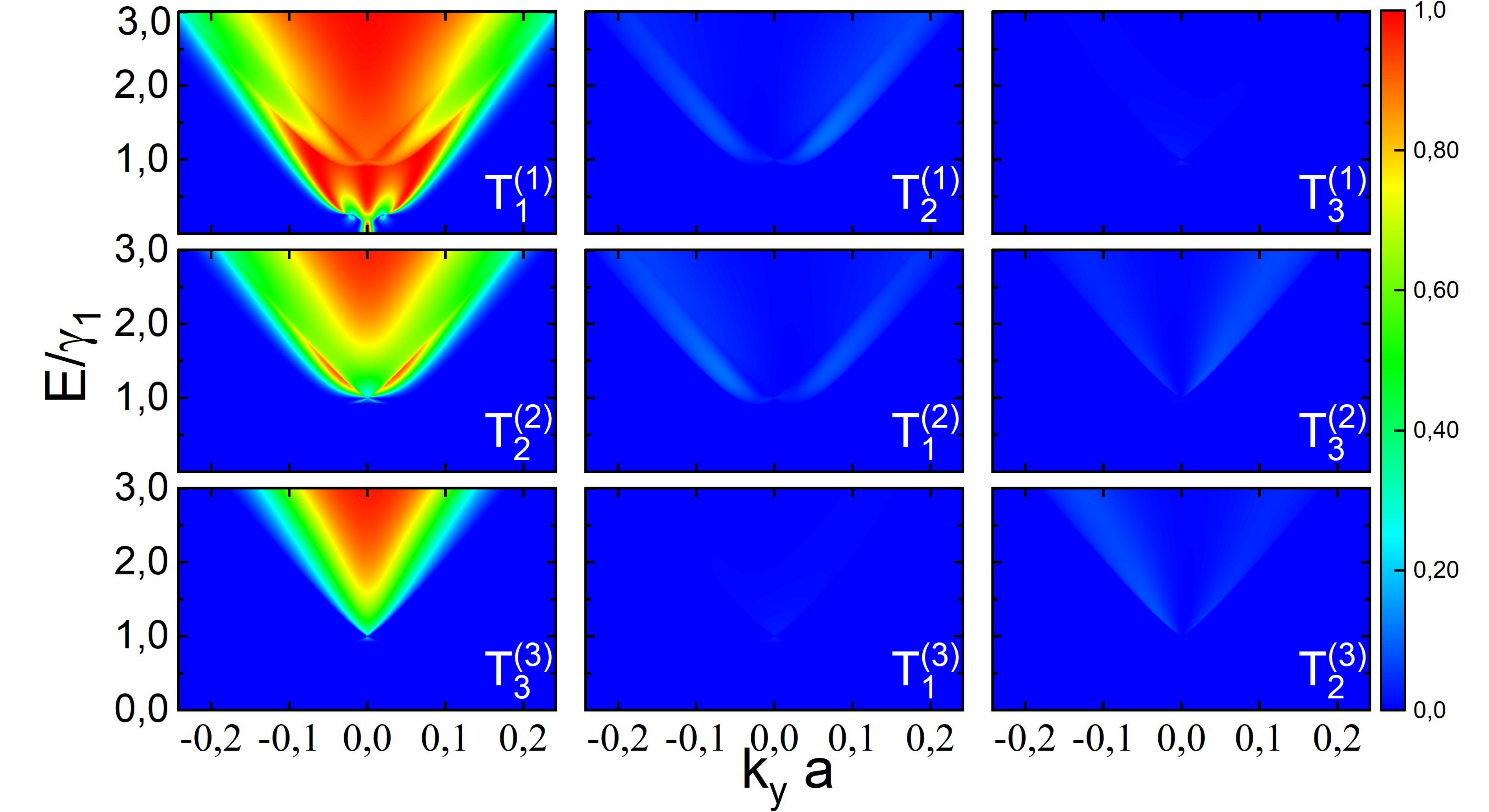}
\caption{(Color online) The same parameters as in Fig. \ref{DensityPlotN=1.pdf} but now for $N=5$.}
\label{DensityPlotN=5.pdf}
\end{figure}

\begin{figure}[tbh]
\begin{center}
\end{center}
\includegraphics[width=\linewidth]{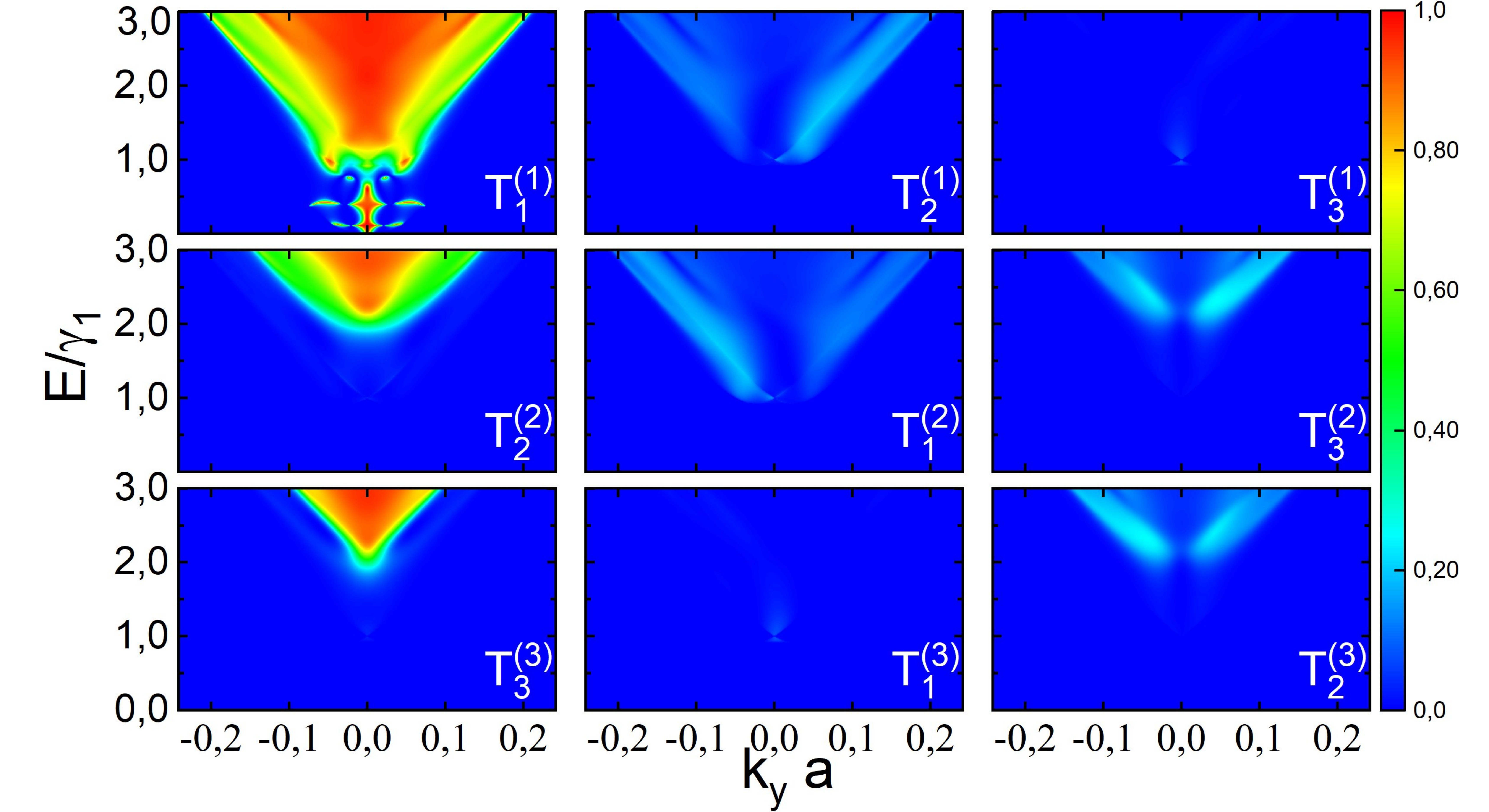}
\caption{(Color online) The same parameters as in Fig. \ref{DensityPlotN=1.pdf} but now for $N=20$.}
\label{DensityPlotN=20.pdf}
\end{figure}

In Figs. \ref{DensityPlotN=20.pdf} and \ref{DensityPlotN=60.pdf}, we show the transmission using the same parameters as in Fig. \ref{DensityPlotN=1.pdf}, but for numbers of cells $N=20$ and $N=60$, respectively. In Fig. \ref{DensityPlotN=20.pdf}, where $N =20$, we observe a substantial decrease in Klein tunneling within the $T^{(1)}_{1}$ channel, at both normal and non-normal incidence. In addition, we observe the formation of a pseudo-gap. When the number of cells is increased at $N=60$ in Fig. \ref{DensityPlotN=60.pdf}, the transmission decreases even more and a gap is obtained, as in ref. \cite{Xu}. On the other hand, in both $T^{(2)}_{2}$ and $T^{(3)}_{3}$ channels, the transmission becomes zero in the energy region beneath $U$ when the number of cells increases, as shown in Figs. \ref{DensityPlotN=20.pdf} and \ref{DensityPlotN=60.pdf}. This is contrary to the results in Figs. \ref{DensityPlotN=1.pdf} and \ref{DensityPlotN=5.pdf} as well as the findings in refs. \cite{Ben,El Mouhafid}. Note also that in all the non-scattered channels, when we compare the results of Fig. \ref{DensityPlotN=20.pdf} with those of Fig. \ref{DensityPlotN=60.pdf}, we notice that the transmission begins to oscillate slightly when the number of cells increases. In the scattered channels, such as $T^{(1)}_{2}(k_y)=T^{(2)}_{1}(-k_y)$ and $T^{(2)}_{3}(k_y)=T^{(3)}_{2}(-k_y)$, the transmission becomes more pronounced as the number of cells increases, see Fig. \ref{DensityPlotN=60.pdf}. However, it is still lower than the results in refs. \cite{Ben,El Mouhafid}. The other scattered channels, $T^{(1)}_{3}$ and $T^{(3)}_{1}$ remain zero and seem to not change when the number of cells changes.


\begin{figure}[tbh]
\begin{center}
\end{center}
\includegraphics[width=\linewidth]{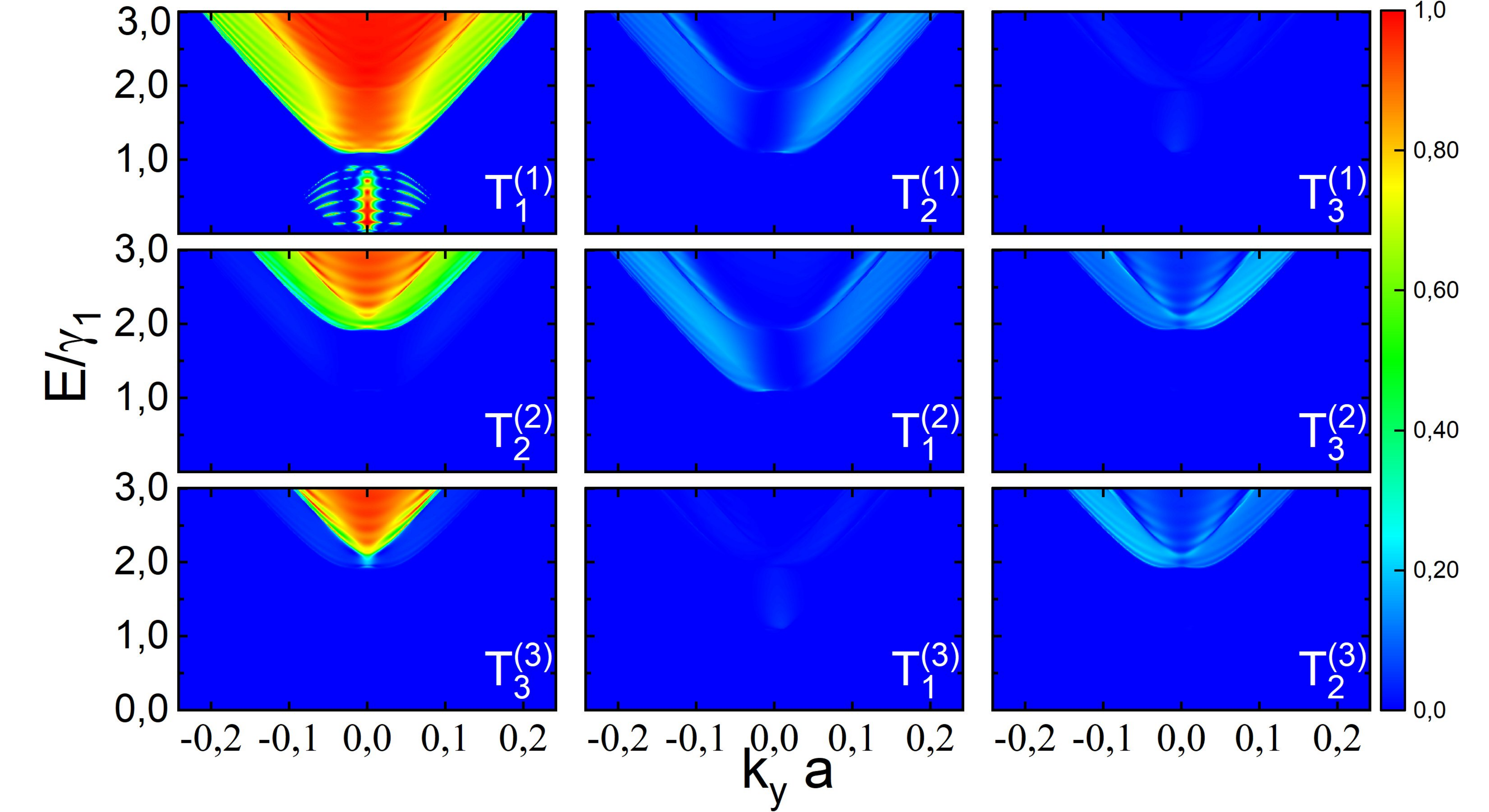}
\caption{(Color online) The same parameters as in Fig. \ref{DensityPlotN=1.pdf} but now for $N=60$.}
\label{DensityPlotN=60.pdf}
\end{figure}

With the same parameters as in Fig. \ref{DensityPlotN=60.pdf}, we plot the transmission in Fig. \ref{DensityPlotN=60_U=2.5.pdf} for barrier height $U=2.5\gamma_1$. As a result, increasing the barrier height decreases the transmission in the non-scattered channels. Indeed, when we compare the channel $T^{(1)}_{1}$ with the result in Fig. \ref{DensityPlotN=60.pdf}, we observe an upward shift of the gap and an increase in anti-Klein tunneling, similar to the result in ref. \cite{Nadia}. In channels $T^{(2)}_{2}$ and $T^{(3)}_{3}$, transmission occurs from $E=2\gamma_1$ in Fig. \ref{DensityPlotN=60.pdf}, where in this case it occurs from $E=2\gamma_1+\delta$, which results in an increase in anti-Klein tunneling. In addition, Klein tunneling is narrowed towards the center in all the non-scattered channels. On the other hand, increasing the barrier height slightly increases the transmission probability in the scattered channels. Particularly, there is a scattering between the modes $k_1$ and $k_3$, which results in a weak transmission in channels $T^{(1)}_{3}(k_y)=T^{(3)}_{1}(-k_y)$, as in ref.  \cite{El Mouhafid}.

In Fig. \ref{DensityPlotN=5_d=3.pdf}, we present the non-scattered channels using the same parameters as those shown in Fig. \ref{DensityPlotN=5.pdf} (where $N=5$), but with a barrier/well width of $d_B = d_W = 3$nm. Similar to the two-band tunneling case, an increase in the barrier/well width results in reduced transmission in all channels. As a consequence, we have two gaps and pseudo-gaps that occur in channel $T^{(1)}_{1}$. On the other hand, channel $T^{(2)}_{2}$ shows three gaps of different sizes, along with one pseudo-gap. In channel $T^{(3)}_{3}$ , two gaps are observed. However, as in Fig. \ref{DensityPlotN=5.pdf}, due to the non-large number of cells, non-zero transmission is found for $\gamma_1<E<U$ in both channels $T^{(2)}_{2}$ and $T^{(3)}_{3}$ in contrast to the results in Figs. \ref{DensityPlotN=20.pdf} and \ref{DensityPlotN=60.pdf}.

\begin{figure}[tbh]
\begin{center}
\end{center}
\includegraphics[width=\linewidth]{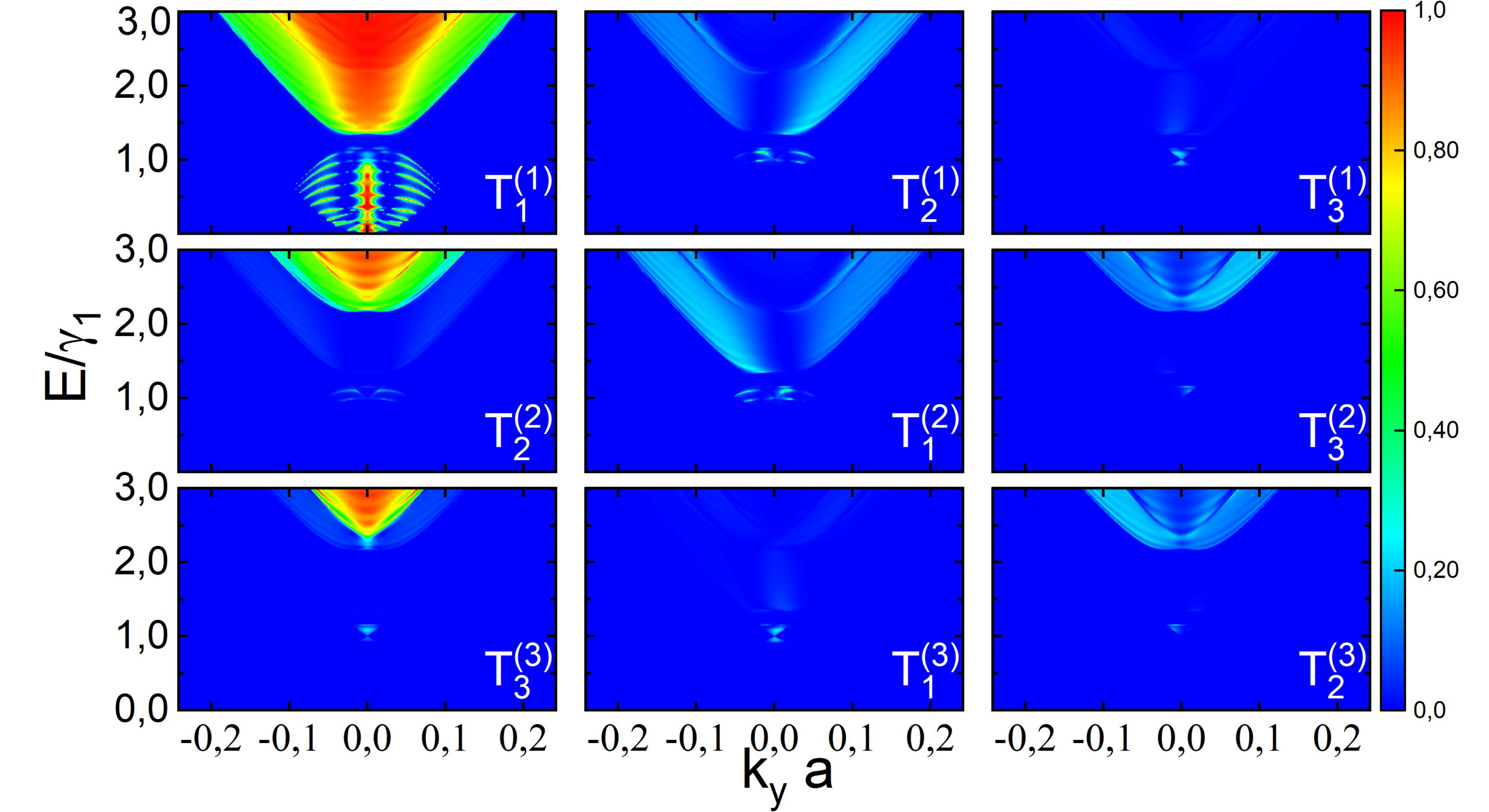}
\caption{(Color online) The same parameters as in Fig. \ref{DensityPlotN=60.pdf} but now for $U=2.5\gamma_1$.}
\label{DensityPlotN=60_U=2.5.pdf}
\end{figure}

\begin{figure}[tbh]
\begin{center}
\end{center}
\includegraphics[width=\linewidth]{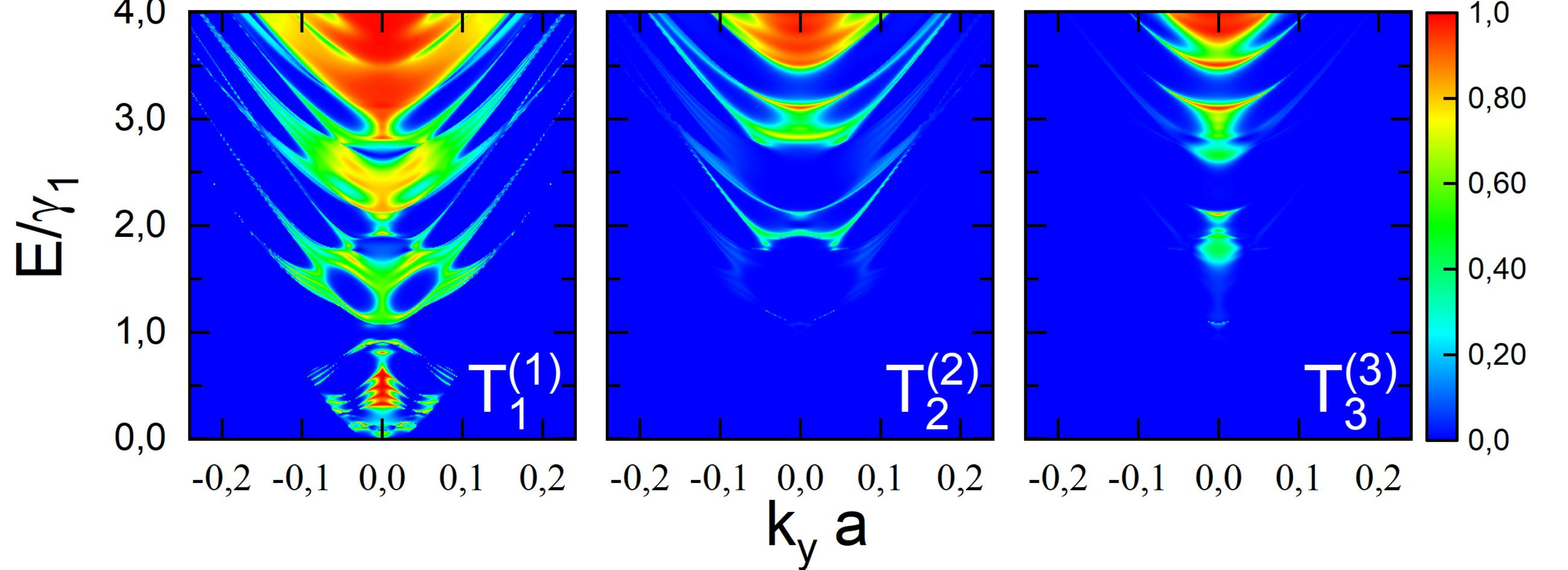}
\caption{(Color online) The same parameters as in Fig. \ref{DensityPlotN=5.pdf} but now for $d_B=d_W=3$nm.}
\label{DensityPlotN=5_d=3.pdf}
\end{figure}

\begin{figure}[tbh]
\begin{center}
\end{center}
\includegraphics[width=\linewidth]{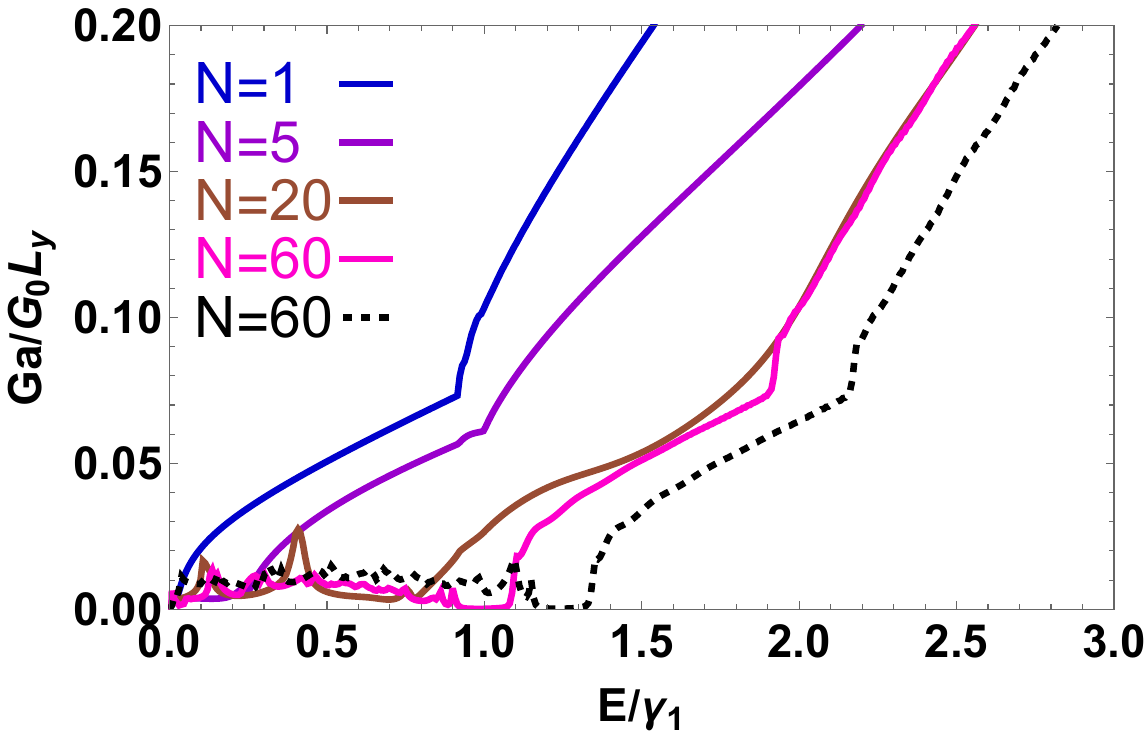}
\caption{(Color online) (solid lines) Total conductance for $N=1$, $5$, $20$ and $60$  corresponding to the transmission in Figs. \ref{DensityPlotN=1.pdf}, \ref{DensityPlotN=5.pdf}, \ref{DensityPlotN=20.pdf} and \ref{DensityPlotN=60.pdf}, respectively. (dashed line) Total conductance for $N=60$ corresponding to the transmission in Fig. \ref{DensityPlotN=60_U=2.5.pdf}.}
\label{conductance}
\end{figure}
Fig. \ref{conductance} displays the total conductance  for $N=1$ (blue line), $N=5$ (purple line) , $N=20$ (brown line) and $N=60$ (magenta line) corresponding respectively to the transmission in Fig. \ref{DensityPlotN=1.pdf}, \ref{DensityPlotN=5.pdf}, \ref{DensityPlotN=20.pdf} and \ref{DensityPlotN=60.pdf}. The black dashed line shows the conductance associated to transmission in Fig. \ref{DensityPlotN=60_U=2.5.pdf} for $N=60$ and barrier height $U=2.5\gamma_1$. For $N=1$, we have high conductance in the region $0<E<\gamma_1$, which results from the full transmission observed in Fig. \ref{DensityPlotN=1.pdf} in channel $T^{(1)}_{1}$ due to the small barrier width\cite{Xu}. When $E>\gamma_1$, there are contribution from $T^{(2)}_{2}$ and $T^{(3)}_{3}$ due to presence of propagation modes $k_2$ and $k_3$ in this region. This increases significantly the conductance. As a result, even though there is no contribution from the scattered channels $T^{\tau}_{i}$ with $\tau\neq i$ (zero transmission), the conductance is higher than the results obtained in refs. \cite{Ben,El Mouhafid}. This is a consequence of the small barrier width that we considered as discussed in Fig. \ref{DensityPlotN=1.pdf}. When the number of cells is increased at $N=5$, the conductance decreases and moves to right side. From number of cells $N=20$, the behaviour of the conductance changes and we observe some minima as in ref \cite{Ben,El Mouhafid,Hassane,Nadia,Saley}. Especially, for $N=60$ a zero conductance is found, which correspond to a gap region (magenta line), in contrast to the result in \cite{El Mouhafid}. {Alternatively, if we increase the barrier height at $U=2.5\gamma_1$ (black dashed line) while maintaining the number of cells at $N=60$, we observe a shift of the gap region to the right. In addition, the conductance decreases as the barrier potential increases, as observed experimentally in \cite{Dubey}. Furthermore, in the region $E<\gamma_1$, the number of oscillations increases with increasing barrier height, which is also consistent with the finding in \cite{Dubey}.}


\section{Conclusion}\label{CC}
In this study, we have investigated the transport properties of Dirac fermions through {ABC-TLG} superlattice. Using the Hamiltonian of the system, we have computed the corresponding eigenvectors and eigenvalues. Subsequently, employing the transfer matrix method and considering the continuity conditions of the spinor at the different interfaces, we have calculated the transmission probability and the corresponding conductance. In the context of two-band tunneling, we observe Klein tunneling at both normal and non-normal incidence for $N=1$. However, as the number of cells increases, Klein tunneling decreases. In addition, oscillations are found at non-normal incidence, and their numbers increase with the number of cells. In the presence of an interlayer bias, a pseudo-gap is found for $N=1$ and turns into a gap when the number of cells is increased. Moreover, the number of oscillations and gaps increases with the barrier/well width.

In the case of six-band tunneling and for non-scattered channels, increasing the number of cells leads to a decrease in the transmission, and oscillations occur. A pseudo-gap is found for a low number of cells, whereas a gap is obtained when the number of cells is increased. It has also been shown that increasing the barrier height at $U=2.5\gamma_1$ decreases the transmission and shifts the gap. On the other hand, in the scattered channels, the transmission is sensitive to the number of cells and the barrier height. Especially in channels $T^{(1)}_{3}$ and $T^{(3)}_1$, the transmission is zero for $U=2\gamma_1$, whereas a weak transmission occurs for $U=2.5\gamma_1$. Finally, it was shown that, increasing the number of cells reduces the conductance such that a gap is found for $N=60$. Our results demonstrate that electron tunneling and gaps can be controlled by adjusting the barrier/well width, the barrier height, and the number of cells. We hope that these findings can serve as a valuable reference for the development of electronic devices using graphene materials.

\section*{Author Contributions}

All authors contributed equally to this work. All authors have read and
approved the published version of the manuscript.

\section*{Declaration of competing interest}

The authors declare that they have no known competing financial interests or
personal relationships that might appear to influence the work presented in
this paper.

\section*{Data Availability Statement}
 This manuscript has no
associated data or the data will not be deposited. [Authors’
comment: The data that support the findings of this study
are available on request from the corresponding author].

\end{document}